\documentclass[12pt]{article}

\usepackage{mytex}
\usepackage{eepic}

\begin{document}

\title{Boundary conditions control for a Shallow-Water model.}
\author{ Eugene Kazantsev
\thanks{
 INRIA, projet MOISE, 
 Laboratoire Jean Kuntzmann,
BP 53,
38041 Grenoble Cedex 9, 
France  }
}
\maketitle

\begin{abstract}
A variational data assimilation technique was  used  to estimate   optimal discretization of interpolation operators and  derivatives  in the nodes adjacent to the rigid boundary.  Assimilation of  artificially generated observational data in the shallow-water model in a square box and  assimilation of real observations in  the model of the Black sea   are discussed. 
It is shown in both experiments that controlling the discretization of operators near a rigid  boundary can bring the model solution closer to observations as in the assimilation window and beyond the window. This type of control allows also to improve climatic variability of the model.

\end{abstract}




\section{Introduction}

Variational data assimilation technique, first proposed in \cite{Ledimet82}, \cite{ldt86}, is based on the optimal control methods \cite{Lions68} and perturbations theory   \cite{Marchuk75}. This technique allows us to retrieve an optimal data for a given model from heterogeneous observation fields. Since the early 1990's, many mathematical and geophysical teams are involved in the development of the data assimilation strategy. One can cite many papers devoted to this problem, as  in the domain of development of different methods for the  data assimilation  and also in the domain of its applications to the atmosphere and oceans.  

In the beginning, data assimilation methods were intended to identify and reconstruct an optimal initial state for the model. However, the idea that other model's parameters should  also be identified by data assimilation has also been studied and discussed in numerous papers.  
One can cite several examples of using  data assimilation  to identify the bottom topography of simple models (\cite{LoschWunsch}, \cite{Heemink}, \cite{assimtopo}), to control open boundary conditions in coastal and regional models (\cite{shulman97}, \cite{shulman98}, \cite{Taillandier}, \cite{Brummelhuis}) and to determine  other parameters of a model (\cite{zou}, \cite{panchang}, \cite{chertok}) . 
 
Together with these parameters, it seems  to be interesting also to control  boundary conditions on rigid boundaries.
Although the boundary configuration of the ocean is steady and can be measured with much better accuracy than the model's initial state, it is not obvious how to represent it  on the model's grid because of  limited resolution. The coastal line of continents possesses a very fine structure and can only be roughly approximated by the model's grid. Consequently,  boundary conditions  are defined at  the model grid's points which are different from the coast. Even the most evident impermeability condition being placed at a wrong point may lead to some error in the model's solution. If it can improve the model solution,  we may accept the flux  can cross the boundary in places where the boundary is in water, prescribing some integral properties on the flux.
  
Ocean models frequently include strong and thin  boundary currents with intense velocity gradients.  In this case, particular attention must be paid to the approximation of the   boundary layer because  a wrong representation of these currents may be responsible for  drastic deformations of the  global solution (see, for example \cite{VerronBlayo}). This may lead us to control the discretization of the model's operators in the whole boundary layer rather than  only at nodes directly    adjacent to the  boundary. 

Alternative method that is frequently used in geophysical models consists in the grid refinement in zones where boundary layers might occur. However, this implies additional computational cost on each model run. Variational data assimilation may help us to determine the parametrization of the boundary layer once for all model runs and to save the computer time.

Boundary conditions on the rigid boundaries have been controlled by data assimilation for heat equation (see, for example, \cite{ChenLin}, \cite{GillijnsDeMoor}), but this control concerns the linear parabolic  diffusion operator rather than hyperbolic transport and advection  operators that are more important in geophysical models.

 Studies on the possibility to control boundary conditions on rigid boundaries for equations containing hyperbolic operators can be found in  \cite{fxld-mo} on the example of non-linear balance equation, in  \cite{assimbc1} on the example of the wave equation, in \cite{sw-lin} on the example of a linear shallow water model in a square box  and in  \cite{Leredde} and \cite{Lellouche} on the example of the Burgers equation.   The principal possibility to improve the model's solution controlling its boundary values are shown in all these papers. However, as it has been noted in  \cite{Leredde}, particular attention must be paid to the discretization process which must respect several rules. 
  It is the discretization of the model's operators  that takes into account the set of boundary conditions and introduces them into the model. Consequently, instead of controlling boundary conditions themselves, there has been proposed in \cite{sw-lin}  to  identify optimal discretization of differential operators at points adjacent to  boundaries.  This allows  us to control directly the way the boundary conditions influence the model and to control boundary parameters in a more general way.  Boundary conditions participate  in discretized operators, but considering the discretization itself, we take into account additional parameters like the position of the boundary,  lack of resolution of the grid, etc.   In  \cite{assimbc1}, for example, it was shown that deplacing  the boundary helps to correct numerical error resulting in   a wrong wave velocity and this displacement has been directly introduced in the discretization of derivatives.

In this paper, we apply 4D-Var  data assimilation to control the discretization of derivatives and interpolation operators in the boundary regions of a non-linear shallow-water model. We use methods proposed in \cite{sw-lin} and study the assimilation results obtained with  a model that exhibits a chaotic behavior.   Two examples are considered in this paper: an academic case of the model in a  square box with artificially generated observations and  more realistic case of  assimilation of   real observational data in the Black sea model.

\section{Shallow Water Model}

\subsection{Model's equations and discretization}
\label{sec1}

The data are assimilated into  the  shallow-water model on the $\beta$-plane \cite{Gill}, \cite{Pedlosky}:
\beqr
\der{u}{t} -(  f_0+\beta y -\der{u}{y})v+ \der{(u^2/2+g\eta)}{x}  &=&  -\sigma u +\mu\Delta u +\fr{\tau_x}{\rho_0 H_0}, 
\nonumber \\
\der{v}{t} +(f_0+\beta y +\der{v}{x})u+ \der{(v^2/2+g\eta)}{y}&=&  -\sigma v +\mu\Delta v +\fr{\tau_y}{\rho_0 H_0}, 
\label{sw}  \\
\der{\eta}{t} + \der{\eta u}{x}+\der{\eta v}{y}&=&0. \nonumber
\eeqr
where $u(x,y,t)$ and $v(x,y,t)$ are  two velocity components, $\eta(x,y,t)$ is the sea surface elevation, $\rho_0$ is the mean density of water, $H_0$ is the characteristic depth of the basin and  $g$ is the reduced gravity. The model is driven by the surface wind stress with components $\tau_x(x,y)$ and $\tau_y(x,y)$ and subjected to the  bottom drag that is parametrized by  linear terms $\sigma u,\; \sigma v$. Horizontal eddy diffusion is represented by   harmonic operators  $\mu\Delta u$ and $\mu\Delta v$. Coriolis parameter is supposed to be linear in $y$ coordinate and is presented as $(f_0+\beta y)$.  The system is defined in some domain $\Omega$ with characteristic size $L$  requiring that   both  $u$ and $v$ vanish on the whole boundary of $\Omega$. No boundary conditions is prescribed for $\eta$.  Initial conditions are defined for all variables $u,\; v$ and $\eta$. 

We discretize all variables of this equation on the regular  Arakawa's C-grid \cite{AL77} with constant grid step $h=\fr{L}{N}$ in both $x$ and $y$ directions (see \rfg{grid})
\beqr
u_{i,j-1/2}(t)&=&u(i h,j h-h/2,t) \mbox{ for } i=0,\ldots N, j=0,\ldots,N+1 \nonumber \\
v_{i-1/2,j}(t)&=&v(ih-h/2,jh,t) \mbox{ for } i=0,\ldots N+1, j=0,\ldots N \nonumber \\
\eta_{i-1/2,j-1/2}(t)&=&\eta(ih-h/2,jh-h/2,t) \mbox{ for } i=0,\ldots N+1, j=0,\ldots N+1 \nonumber 
\eeqr

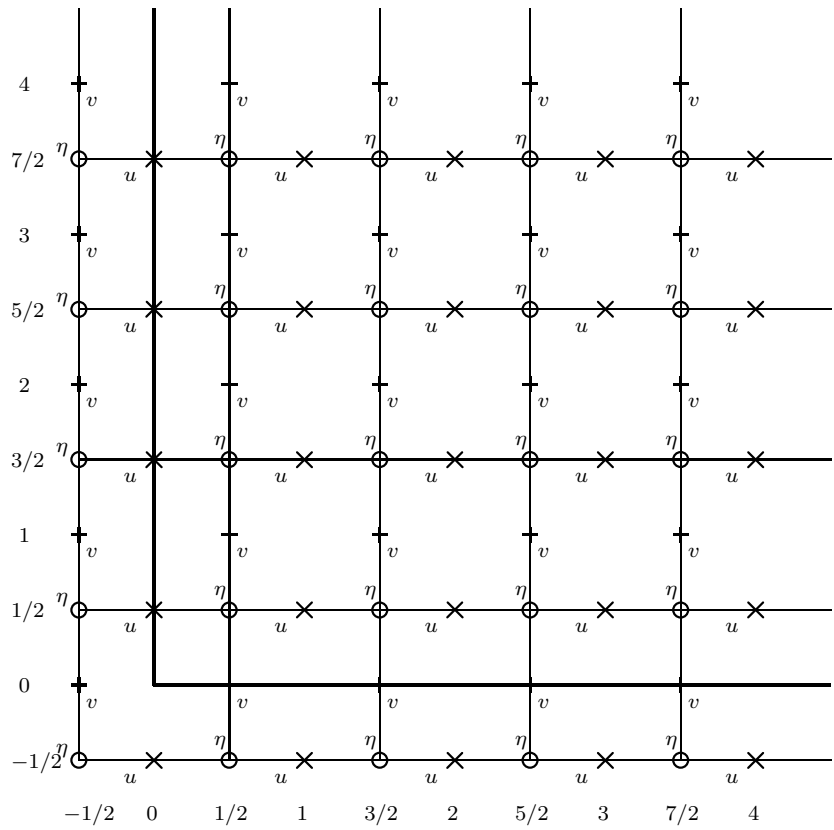
\begin{figure}
\setlength{\unitlength}{1mm}
\newcount\indi
\newcount\indj
\newcount\num
\begin{center}
\begin{picture}(100,120)
\scriptsize
\multiput(10,10)(0,20){5}{\line(1,0){100}}
\multiput(10,10)(20,0){5}{\line(0,1){100}}
\Thicklines
\put(20,20){\line(1,0){90}}
\put(20,20){\line(0,1){90}}
\thicklines
\multiput(19,9)(20,0){5}{\line(1,1){2}}
\multiput(21,9)(20,0){5}{\line(-1,1){2}}
\indi=-1
\multiput(15,7)(20,0){5}{ 
\global\advance\indi by 1 $u$ }

\multiput(19,29)(20,0){5}{\line(1,1){2}}
\multiput(21,29)(20,0){5}{\line(-1,1){2}}
\indi=-1
\multiput(15,27)(20,0){5}{ 
\global\advance\indi by 1 $u$ }

\multiput(19,49)(20,0){5}{\line(1,1){2}}
\multiput(21,49)(20,0){5}{\line(-1,1){2}}
\indi=-1
\multiput(15,47)(20,0){5}{ 
\global\advance\indi by 1 $u$ }

\multiput(19,69)(20,0){5}{\line(1,1){2}}
\multiput(21,69)(20,0){5}{\line(-1,1){2}}
\indi=-1
\multiput(15,67)(20,0){5}{ 
\global\advance\indi by 1 $u$ }

\multiput(19,89)(20,0){5}{\line(1,1){2}}
\multiput(21,89)(20,0){5}{\line(-1,1){2}}
\indi=-1
\multiput(15,87)(20,0){5}{ 
\global\advance\indi by 1 $u$ }
\multiput(9,20)(0,20){5}{\line(1,0){2}}
\multiput(10,19)(0,20){5}{\line(0,1){2}}
\indj=-1
\multiput(10,17)(0,20){5}{ 
\global\advance\indj by 1 $v$ }

\multiput(29,20)(0,20){5}{\line(1,0){2}}
\multiput(30,19)(0,20){5}{\line(0,1){2}}
\indj=-1
\multiput(30,17)(0,20){5}{ 
\global\advance\indj by 1 $v$ }

\multiput(49,20)(0,20){5}{\line(1,0){2}}
\multiput(50,19)(0,20){5}{\line(0,1){2}}
\indj=-1
\multiput(50,17)(0,20){5}{ 
\global\advance\indj by 1 $v$ }

\multiput(69,20)(0,20){5}{\line(1,0){2}}
\multiput(70,19)(0,20){5}{\line(0,1){2}}
\indj=-1
\multiput(70,17)(0,20){5}{ 
\global\advance\indj by 1 $v$ }

\multiput(89,20)(0,20){5}{\line(1,0){2}}
\multiput(90,19)(0,20){5}{\line(0,1){2}}
\indj=-1
\multiput(90,17)(0,20){5}{ 
\global\advance\indj by 1 $v$ }

\multiput(10,10)(0,20){5}{\circle{2}}
\indj=-1
\multiput(6,11)(0,20){5}{ 
\global\advance\indj by 1 \num=\indj \multiply\num by 2 \global\advance\num by -1 $\eta$ }

\multiput(30,10)(0,20){5}{\circle{2}}
\indj=-1
\multiput(27,12)(0,20){5}{ 
\global\advance\indj by 1 \num=\indj \multiply\num by 2 \global\advance\num by -1 $\eta$ }

\multiput(50,10)(0,20){5}{\circle{2}}
\indj=-1
\multiput(47,12)(0,20){5}{ 
\global\advance\indj by 1 \num=\indj \multiply\num by 2 \global\advance\num by -1 $\eta$ }

\multiput(70,10)(0,20){5}{\circle{2}}
\indj=-1
\multiput(67,12)(0,20){5}{ 
\global\advance\indj by 1 \num=\indj \multiply\num by 2 \global\advance\num by -1 $\eta$ }

\multiput(90,10)(0,20){5}{\circle{2}}
\indj=-1
\multiput(87,12)(0,20){5}{ 
\global\advance\indj by 1 \num=\indj \multiply\num by 2 \global\advance\num by -1 $\eta$ }

\indj=-1
\multiput(7,2)(20,0){5}{ 
\global\advance\indj by 1 \num=\indj \multiply\num by 2 \global\advance\num by -1 $\the\num/2$ }
\indi=-1
\multiput(18,2)(20,0){5}{ 
\global\advance\indi by 1 $\the\indi$ }

\indj=-1
\multiput(0,9)(0,20){5}{ 
\global\advance\indj by 1 \num=\indj \multiply\num by 2 \global\advance\num by -1 $\the\num/2$ }

\indi=-1
\multiput(1,19)(0,20){5}{ 
\global\advance\indi by 1 $\the\indi$ }

\end{picture} 
\end{center}
\refstepcounter{fig}
\label{grid}
\caption{Arakawa C-grid}
\end{figure}

In order to discretize the system \rf{sw}, we replace the derivatives by their finite difference representations  $D_x$ and $D_y$ and   introduce  two    interpolations in $x$ and $y$ coordinates $S_x$ and $S_y$. Interpolations are necessary on the staggered grid  to calculate the variable's values in nodes where other variables are defined.  Following \cite{AL81}, we write   the discretized system \rf{sw} as
\beqr
\der{u}{t} -(  f_0+\beta y -S_y D_y u)S_x S_y v+ D_x((S_x u)^2/2+g\eta)  &=&  -\sigma u +\mu\Delta^h u +\fr{\tau_x}{\rho_0 H_0}, 
\nonumber \\
\der{v}{t} +(f_0+\beta y +S_x D_x v)S_y S_x u+ D_y((S_y v)^2/2+g\eta)&=&  -\sigma v +\mu\Delta^h v +\fr{\tau_y}{\rho_0 H_0}, 
\label{sw-grid}  \\
\der{\eta}{t} + D_x(uS_x\eta )+D_y(v S_y\eta  )&=&0. \nonumber
\eeqr

Discretized operators $D_x, D_y$ and $S_x, S_y $ are defined in a classical way at all internal points of the domain. For example,  the derivative and the interpolation operator of the variable $u$ defined at corresponding  points  write
\beqr
(D_x u)_{i-1/2,j-1/2}&=&\fr{u_{i,j-1/2} -u_{i-1,j-1/2} }{h}  \mbox{ for } i=2,\ldots , N-1,
 \nonumber \\
 \label{intrnlsch} \\
(S_x u)_{i-1/2,j-1/2}&=&\fr{u_{i,j-1/2} +u_{i-1,j-1/2} }{2}\mbox{ for } i=2,\ldots , N-1. \nonumber
\eeqr
These expressions  represent a well known second order approximation. 
Laplace operator is discretized in a common way $\Delta^h v = \fr{v_{i+1,j}+v_{i-1,j}-2v_{i,j}}{h^2}+\fr{v_{i,j+1}+v_{i,j-1}-2v_{i,j}}{h^2}.$

The boundary region  is considered as a band  of adjacent to boundary nodes.   Discretizations of operators in this band are  different from \rf{intrnlsch} and represent the control variables in this study. In order to obtain their  optimal values  assimilating external data, we suppose nothing about derivatives near the boundary  and  write them in a general form
\beq
(D_x u)_{1/2,j-1/2}=\fr{\alpha_{0}^{D_xu}+\alpha^{D_xu}_{1} u_{0,j-1/2} +\alpha^{D_xu}_{2} u_{1,j-1/2}}{h}  \label{bndsch}
\eeq
This formula represents a linear combination of values of $u$ at two points adjacent to the boundary with  coefficients $\alpha$. These coefficients  are considered as particular for each operator and for each variable. The constant $\alpha_0$, which is  also particular  for each operator, may be added in some cases to simulate non uniform boundary conditions like $u(0,y)=\alpha_0^{D_xu}\neq 0$.

We distinguish $\alpha$ for different variables and different operators allowing different controls of  derivatives   because of the different nature of these variables and different boundary conditions prescribed for them. It is obvious, for example, that the approximation of the derivative of $\eta$ in the gradient may differ from the approximation of the derivative of $u$ in the divergence. Although both operators represent a derivative,  boundary conditions for $u$ and $\eta$ are different, these derivatives are defined at different points, at different distance from the boundary. Consequently, it is reasonable to let them be controlled separately and  to assume that their optimal approximation may be different with distinct coefficients $\alpha^{D_xu}$ and $\alpha^{D_x\eta}$.

Time stepping of this model is performed by the leap-frog scheme for all hyperbolic terms and Euler scheme for the dissipative terms. 
The first time step is splitted into two Runge-Kutta stages in order to ensure the second order approximation.

\subsection{Tangent and adjoint models}

The approximation of the derivative introduced by \rf{intrnlsch} and \rf{bndsch} depends on control variables $\alpha$. Operators  are allowed to change their properties near boundaries in order to find the best fit with requirements of the model and data.   To assign  variables $\alpha$ we shall perform data assimilation procedure and find their optimal values.  Variational data assimilation is usually performed by minimization of the specially introduced cost function. The minimization is achieved using the gradient of the cost function that is usually determined by the run of the adjoint to the tangent linear model.

Developing the  tangent and adjoint models in this case, we follow the procedure presented in \cite{sw-lin}. The tangent  model  is equal to the Gateaux derivative of the original model \rf{sw-grid} with respect to the control parameters. To calculate this derivative we suppose that the control variables  $\alpha$ can have small variations $\delta\alpha$ and we determine the linear model that governs the evolution of the perturbations   $\delta u, \; \delta v, \;\delta \eta$  of the solution $u,\;v,\;\eta$ of the model \rf{sw-grid}. Skipping the detailed development of the tangent model (see \cite{sw-lin}), we write the model as
\scriptsize
\beq
\der{}{t}
\left(\begin{array}{c}
 \delta u\\ \delta v\\ \delta\eta\\ \delta\alpha
\end{array}\right)=
\left(\begin{array}{cccc}
\begin{array}{c}
S_xS_yv\cdot S_yD_y \\ +D_x(S_xu\cdot S_x)
-\sigma+\mu\Delta^h 
\end{array}    &-(f_0+\beta y-S_yD_yu)\cdot S_x S_y & -g D_x&R_u\\
 (f_0+\beta y+S_xD_xv)\cdot  S_y S_x&
\begin{array}{c}
S_yS_xu\cdot S_xD_x\\
 +D_y(S_yv\cdot S_y) -\sigma+\mu\Delta^h 
\end{array}
&-g D_y&R_v\\
 D_x(S_x\eta\cdot )& D_y(S_y\eta\cdot )& 0&R_\eta\\
0&0&0&0
\end{array}\right)
\left(\begin{array}{c}
 \delta u\\ \delta v\\ \delta\eta\\ \delta\alpha
\end{array}\right) \label{tlm}
\eeq
\normalsize
where
\beqr
R_u\delta\alpha&=& -(f_0+\beta y-S_yD_yu)\cdot(\delta S_x (S_y v)+S_x (\delta S_y v)) +S_xS_yv\cdot(\delta S_y(D_yu)+S_y(\delta D_yu))+ \nonumber \\
&+& \delta D_x((S_xu)^2/2+g\eta) +D_x(S_xu\cdot\delta S_xu)  
\nonumber \\
R_v\delta\alpha&=&(f_0+\beta y+S_xD_xv)\cdot(\delta S_y(S_x u)+S_y (\delta S_x u))+S_yS_xu\cdot(\delta S_x(D_xv) +S_x(\delta D_xv)) + \nonumber\\
&+&\delta D_y((S_yv)^2/2+g\eta) +D_y(S_yv\cdot \delta S_yv))
\label{R}\\
R_\eta\delta\alpha&=&\delta D_x(u\cdot S_x\eta) + D_x(u\cdot \delta S_x \eta)+\delta D_y(v\cdot S_y\eta) + D_y(v\cdot \delta S_y \eta) \nonumber
\eeqr
Operators $\delta S$  and $\delta D$ denote $S(\alpha +\delta \alpha)-S(\alpha)$ and $D(\alpha +\delta \alpha)-D(\alpha)$ respectively. These operators are  of  implicit structure. Their argument (that is, in fact,  $\delta\alpha$)  is contained in the matrix itself. This representation is not convenient in the development of  the adjoint model, that's why we would better rewrite them in a more classical form: a constant operator (which does not depend on $\delta\alpha$)   multiplied by a variable  vector (which is $\delta\alpha$). So, each product like $\delta D_x u,\; \delta S_x \eta$ etc. is replaced by another product:
$\delta D_x u=\widehat{[u]}_{D_x} \vec\delta\alpha,\; \delta S_x \eta=\widehat{[\eta]}_{S_x}\vec\delta\alpha $ etc.
where   operators $\widehat{[u]},\; \widehat{[\eta]}, \ldots $ are constructed from the solution $u, \eta$  of the original equation  and the vector $\vec{\delta\alpha}$ is extracted from  matrices $\delta D$ or $\delta S$ as it is described in \cite{assimbc1}. We shall further use hats to denote matrices that have been constructed from vectors. These matrices are also block-matrices. All elements of their blocks are equal to zero except one  line in the beginning and one line at the end  of each block. These lines are composed of two values of corresponding vector, and, namely, values of approximated function in two nodes  near the boundary.

Using these notations, operators $R$ \rf{R} are rewritten as   
\beqr
R_u&=& -(f_0+\beta y-S_yD_yu)\cdot(\widehat{[S_y v]}_{S_x}+S_x \widehat{[v]}_{S_y}) +S_xS_yv\cdot( \widehat{[D_yu]}_{S_y}+S_y\widehat{[u]}_{D_y})+ \nonumber \\
&+& \widehat{[(S_xu)^2/2+g\eta]}_{D_x} +D_x(S_xu\cdot\widehat{[u]}_{S_x})  
\nonumber \\
R_v&=&(f_0+\beta y+S_xD_xv)\cdot(\widehat{[S_x u]}_{S_y}+S_y \widehat{[u]}_{S_x}) +S_yS_xu\cdot( \widehat{[D_xv]}_{S_x} +S_x\widehat{[v]}_{D_x}) + \nonumber\\
&+&\widehat{[(S_yv)^2/2+g\eta]}_{D_y} +D_y(S_yv\cdot \widehat{[v]}_{S_y})
\label{R-mod}\\
R_\eta&=&\widehat{[u\cdot S_x\eta]}_{D_x} + D_x(u\cdot \widehat{[\eta]}_{S_x})+\widehat{[v\cdot S_y\eta]}_{D_y} + D_y(v\cdot \widehat{[\eta]}_{S_y}) \nonumber
\eeqr
  
The tangent model is subjected to  the same zero boundary conditions  for $\delta u$ and $\delta v$  as for $u$ and $v$. No boundary conditions is prescribed for  $\delta\eta$ as well as for $\eta$. At initial time   $\delta u, \delta v $ and $ \delta \eta$ are equal to  $\delta u_0, \delta v_0 $ and $ \delta \eta_0$ respectively. These variables are now classical in controlling initial conditions of the model. They will be used for the same purpose and for the joint control of a boundary scheme and initial conditions of the model as well.  

We should note,  the matrix of the  tangent linear model \rf{tlm} is  composed by two  parts: the $3\tm3$ block composed of operators acting in the space of the model's variables and the fourth column composed of operators $R$ \rf{R-mod}.  The $3\tm3$ block is responsible for the evolution of a small perturbation by the model's dynamics and is similar for any data assimilation.  The column  determines the way how this perturbation is introduced into the model and is specific to the particular variable under control. This column is absent when the goal is to identify the initial conditions of the model because the uncertainty in initial conditions  is introduced only once, at the beginning of the model integration. But, when the uncertainty is presented in an internal model parameter, like in this case, the perturbation is introduced at each time step of  the model.

  In order to develop the adjoint model, we need to introduce the scalar product in the space defined by tangent model. Each element in this space is composed of discretized functions $ u,\; v$ and $\eta$ and also the whole set   of the control coefficients $\alpha$. A vector in this space has four components $\phi=( u, v, \eta, \alpha)$. 

Following \cite{sw-lin}, we consider a weighted  Euclidian scalar product in this space 
\beqr
\spm{\phi}{\phi^*}&=&\spm{\left(\begin{array}{c}
   u\\   v\\  \eta\\  \alpha
\end{array}\right) }{
 \left(\begin{array}{c}
  u^*\\  v^*\\ \eta^*\\ \alpha^*
\end{array}\right)} = \label{sp}\\
&=& w_u^2\sum_{i,j} u_{i,j} u^*_{i,j} + w_v^2\sum_{i,j} v_{i,j} v^*_{i,j}+w_\eta^2\sum_{i,j} \eta_{i,j} \eta^*_{i,j} +\sum_{k,operator} \alpha_{k}^{operator} (\alpha^{operator}_{k})^*
\nonumber
\eeqr
The sums in the scalar product is performed over all nodes $i,j$  of the grid  of all model's variables $u,\;v$ and $\eta$. The sum of control coefficients $\alpha$ is performed over all $k: 0\leq k\leq 2,\; $ \rf{bndsch}  and over all operators ``$operator$'' controlled by these coefficients.
Weights $w_u=w_v=\fr{1}{\sqrt{gH_0}}$ and $w_{\eta}=\fr{1}{H_0}$ are introduced to  bring all variables to dimensionless form. 
Applying the development described in \cite{sw-lin} we get the  formulation of the   adjoint model:
\scriptsize
\beq
-\der{}{t}
\left(\begin{array}{c}
  u^*\\  v^*\\ \eta^*\\ \alpha^*
\end{array}\right)=
\left(\begin{array}{cccc}
\begin{array}{c}
 D_y^*S_y^*(S_xS_yv\cdot )+\\ +S_x^*(S_xu\cdot D_x^* )-\sigma+\mu\Delta^h   
\end{array}
 &S_x^*S_y^*(f_0+\beta y+S_xD_xv) \cdot & S_x\eta\cdot D_x^*&0\\
- S_y^* S_x^* (f_0+\beta y-S_yD_yu)\cdot&
\begin{array}{c}
  D_x^*S_x^*(S_yS_xu\cdot )+\\ +S_y^*(S_yv\cdot D_y^* ) -\sigma+\mu\Delta^h
\end{array}
&S_y\eta\cdot D_y^*&0\\
g D_x^*& g D_y^*& 0&0\\
R_u^*&R_v^*&R_\eta^*&0
\end{array}\right)
\left(\begin{array}{c}
  u^*\\  v^*\\ \eta^*\\ \alpha^*
\end{array}\right) \label{am}
\eeq
\normalsize
where
\beqr
R_u^*&=& -(\widehat{[S_yv]}^*_{S_x}+\widehat{[v]}^*_{S_y}S_x^*) ((f_0+\beta y-S_yD_yu)\cdot )+
(\widehat{[D_yu]}^*_{S_y}+\widehat{[u]}^*_{D_y}S_y^*)(S_xS_yv\cdot )+\nonumber \\
 &+&\widehat{[(S_xu)^2/2+g\eta]}^*_{D_x}  +\widehat{[u]}^*_{S_x}(S_xu\cdot D_x^*)\nonumber \\
R_v^*&=& (\widehat{[S_xu]}^*_{S_y}+\widehat{[u]}^*_{S_x}S_y^*) ((f_0+\beta y-S_xD_xv)\cdot )+
(\widehat{[D_xv]}^*_{S_x}+\widehat{[v]}^*_{D_x}S_x^*)(S_yS_xu\cdot )+\nonumber \\
 &+&\widehat{[(S_yv)^2/2+g\eta]}^*_{D_y}  +\widehat{[v]}^*_{S_y}(S_yv\cdot D_y^*)\label{R*} \\
R_\eta^*&=& \widehat{[u\cdot S_x\eta]}^*_{D_x} +
\widehat{[\eta]}^*_{S_x} (u\cdot D_x^*)+\widehat{[v\cdot S_y\eta]}^*_{D_y} +
\widehat{[\eta]}^*_{S_y} (v\cdot D_y^*)
\nonumber 
\eeqr

\subsection{Cost function}

One of the principal  purposes of variational data assimilation consists in the variation of  control parameters in order to bring the model's solution closer to the observational data. This implies the necessity to measure the distance between the trajectory of the model and data. Introducing the cost function, we define this measure. Generally speaking, the cost function is represented by some norm of the difference between model's solutions and observations, eventually accompanied by some regularization term. 

To characterize the difference between the model's solution and  the observational data we use the norm based on the scalar product  \rf{sp} with the fourth component $\alpha$ equal to zero:
\beq 
\xi^2=w_u^2\sum_{i,j} (u_{i,j}- u^{obs}_{i,j})^2 + w_v^2\sum_{i,j}(v_{i,j}- v^{obs}_{i,j})^2+w_\eta^2\sum_{i,j} (\eta_{i,j}- \eta^{obs}_{i,j})^2. \label{xi}
\eeq
 Expressing $\xi$ in terms of the scalar product, we emphasize its dependence on time and on control coefficients $\alpha$:
\beqr
\xi^2&=&\xi^2(\alpha,t)=\spm{\phi(\alpha,t)-\phi^{obs}(t)}{\phi(\alpha,t)-\phi^{obs}(t)}=
\nonumber\\
&=&\spm{\left(\begin{array}{c}
   u(\alpha,t)-u^{obs}(t)\\   v(\alpha,t)-v^{obs}(t)\\  \eta(\alpha,t)-\eta^{obs}(t)\\  0
\end{array}\right) }{\left(\begin{array}{c}
   u(\alpha,t)-u^{obs}(t)\\   v(\alpha,t)-v^{obs}(t)\\  \eta(\alpha,t)-\eta^{obs}(t)\\  0
\end{array}\right) 
 }
\eeqr
Taking into account the results obtained in \cite{sw-lin}, we define the cost function as 
\beq
\costfun(\alpha)=\int\limits_0^T  t \xi^2(\alpha,t) dt \label{costfn}
\eeq
that gives  bigger  importance to the difference $\xi^2$ at the end of assimilation interval.  

It should be noted here, that this cost function can  only  be used in the case of  assimilation of a perfect artificially generated data. When we assimilate some kind of real data that contains errors of measurements and is defined on a different grid, we should add some regularization term to the cost function (like the distance from the initial guess) and use some more appropriate norm instead of the Euclidean one (see, for example  \cite{UnifNotat} for details).  

 To calculate the gradient of the cost function, we  calculate  first its variation:  
 \beqr
 \delta\costfun&=&\costfun(\alpha+\delta\alpha)-\costfun(\alpha)=\int\limits_0^T t(  \xi^2(\alpha+\delta\alpha,t)-\xi^2(\alpha,t)) dt=
  \nonumber \\
&=&\int\limits_0^T t\biggl(\spm{\phi(\alpha+\delta\alpha,t)-\phi^{obs}(t)}{\phi(\alpha+\delta\alpha,t)-\phi^{obs}(t)}-\nonumber \\
&-&\spm{\phi(\alpha,t)-\phi^{obs}(t)}{\phi(\alpha,t)-\phi^{obs}(t)}\biggr) dt \sim
  2\int\limits_0^T t\spm{\delta\phi(t)}{\phi(\alpha,t)-\phi^{obs}(t)} dt
\eeqr 

As it has been shown in \cite{sw-lin}, the scalar product $\spm{\delta\phi(t)}{\phi(\alpha,t)-\phi^{obs}(t)}  $ is equal to $ \spm{\delta\phi(0)}{{\cal A}^*(t,0)(\phi(\alpha,t)-\phi^{obs}(t))} $  where ${\cal A}^*(t,0) $ denotes the adjoint model \rf{am} integrated from the state  $\phi(\alpha,t)-\phi^{obs}(t)$ back in time from $t$ to $0$.  So, the result of the adjoint model run, being scalarly multiplied by $\delta\phi(0)$ provides the variation of the cost function
\beq
\delta\costfun= 2 \spm{\delta\phi(0)}{\int\limits_0^T t{\cal A}^*(t,0)(\phi(\alpha,t)-\phi^{obs}(t)) dt} 
\eeq
As it has been mentioned above, vector $ \delta\phi(0)$ is composed of 4 components: $\delta u_0, \;\delta v_0, \;\delta \eta_0$  and $\delta\alpha $. If we want to control the boundary scheme only,  we put zero to the first three components of $\delta\phi(0)$. Only the fourth component of  the  variation of the cost function (and its gradient) is different from zero in this case and only this component is used in the control. On the other hand, if our purpose is to control initial state of the model, then vanishing $\delta\alpha $ must be imposed and the first three components of the gradient must be used.  And,  for the joint control of boundary and initial conditions of the model we use all four components of the gradient. 

 Thus, the gradient of the cost function writes
 \beq
 \nabla \costfun =2\int\limits_0^T t{\cal A}^*(t,0) (\phi(\alpha,t)-\phi^{obs}(t)) dt.
 \label{grad}
 \eeq

This gradient is used in the minimization procedure that is implemented in order  to find the minimum of the cost function:
\beqr
&\costfun(\bar\alpha) = \min\limits_{\alpha} \costfun(u_0,v_0,\eta_0,\alpha)& \mbox{
\begin{minipage} [l]{0.4\textwidth} 
if we control the discretization of operators near the boundary  using only the fourth component of the gradient \rf{grad};\end{minipage}
}\label{1stway}\\
&\costfun(\bar u_0, \bar v_0,\bar \eta_0  ) = \min\limits_{u_0,v_0,\eta_0} \costfun(u_0,v_0,\eta_0,\alpha)&  
\mbox{\begin{minipage} [l]{0.4\textwidth} 
 if we control  only the initial state of the model using three  components of the gradient;\end{minipage}}\label{2ndway}\\
&\costfun(\bar u_0, \bar v_0,\bar \eta_0,\bar\alpha  ) = \min\limits_{u_0,v_0,\eta_0,\alpha} \costfun(u_0,v_0,\eta_0,\alpha) &
 \mbox{\begin{minipage} [l]{0.4\textwidth}  if we perform the joint control of both the initial state and the boundary discretization of the model using all four  components of the gradient.\end{minipage}}
\label{3rdway}
\eeqr

Coefficients $\bar\alpha$  are considered as coefficients achieving an  optimal discretization of the model's operators in the boundary regions. 
We use the  minimization procedure  developed by Jean Charles Gilbert and  Claude Lemarechal, INRIA \cite{lemarechal}.  The procedure uses the limited memory quasi-Newton method.

\section{Model in a square box. }

We start from the data assimilation in frames of the very well studied "academic" configuration. 
Several experiments have been performed with the  model  in a square box of side length $L=2000$ km driven by a steady, zonal  wind forcing with a classical sinusoidal profile
$$
\tau_x=\tau_0 \cos \fr{2\pi (y-L/2)}{L}
$$
that leads to the formation of a double gyre circulation \cite{LPV}. The attractor of the model and the bifurcation diagram  in a similar configuration has been described in \cite{simmonet2}. Following their results, we intentionally chose the model's parameters in order to ensure chaotic behavior. The maximal wind tension on the surface is taken to be $\tau_0=0.5\fr{dyne}{cm^2}$.  The coefficient of Eckman dissipation and the  lateral friction coefficient  are chosen as
$\sigma=5\tm 10^{-8}s^{-1}$ and  $\mu=200\fr{m^2}{s}$ respectively. 

As it has been already noted, the Coriolis parameter is a linear function in $y$ with  $f_0=7\tm 10^{-5}s^{-1}$ and $\beta=2\tm 10^{-11} (ms)^{-1}$. The reduced gravity and the depth are respectively equal to $g=0.02\fr{m}{s^2},\;H_0=1000m$. 

All operators in the model are approximated with the second order accuracy both in the interior of the domain and near its boundary. That means the expression \rf{bndsch}, that is used to interpolate functions and to calculate their derivatives near boundary, is written with  $\alpha^{D}_{1}=-1/h,\; \alpha^{D}_{2}=1/h$ for all derivative operators and $\alpha^{S}_{1}=\alpha^{S}_{2}=1/2$ for all interpolations. That gives, for example, the value of the derivative of $u$ at the point $i=1/2$ as $(D_x u)_{1/2,j-1/2}=\fr{u_{1,j-1/2}-u_{0,j-1/2}}{h}   $.

The solution of the model in this configuration possesses a boundary layer near the Western boundary. This is a  well known Munk layer \cite{Munk}  that is characterized by the local equilibrium between the $\beta$-effect and the lateral dissipation. It's width is expressed by the formula  $\fr{2\pi}{\sqrt{3}}(\mu/\beta)^{1/3}=52.3$ km for the given parameters.

 As it has been  discussed in \cite{Bryan75}, the model must resolve this layer with at least one grid point (optimally, more than one grid point) in order to maintain numerical stability.  The work of \cite{Griffies} emphasized the importance of having at least two grid points in the Munk layer in order to minimize the level of spurious oscillations visible in the velocity fields as well in the field of the sea surface elevation. 

The resolution of the model in this section is intentionally chosen to be too coarse to resolve the Munk layer. The model's grid is composed of 30 nodes in each direction, that means the grid-step is equal to 67 km, that is more than  the Munk layer's width. As it can be seen in \rfg{grid}, there is only one grid node in the layer for variables $v$ and $\eta$ and no nodes at all for $u$ variable. 

Artificial ``observational`` data are generated by
the same model with all the same parameters but with 9 times finer resolution  (7.6 km  grid step). The  fine resolution model, having 7 nodes in the Munk layer, resolves  explicitly the layer and must have no spurious oscillations. All nodes of the coarse grid belong to the fine grid, consequently, we do not need to interpolate ''observational'' data to the coarse grid. We just take values in  nodes of the high resolution grid that correspond to nodes on the coarse grid. 
A profile of the meridional velocity component $v$  is shown in \rfg{profile} in order to illustrate effects of  resolved and  under-resolved  Munk layer. These effects are the most  visible in the field of this variable. However,  similar oscillations are also present in the fields of the velocity $u$ and the sea surface height (SSH)  $\eta$. As it is shown in \rfg{grid}, the  point on the boundary $(x=0)$ does not belong to the grid where $v$ variable is defined. However, boundary conditions are prescribed at this point  imposing vanishing $v$. 

If we compare profiles of the velocity $v(x,y)$ plotted in \rfg{profile} at $y=500$ km (one quarter of the side length), we see that the profile obtained using  the  model with fine  resolution is smooth, while  
the profile on the coarse resolution  (solid line in \rfg{profile}) shows  strong spurious oscillations due to the unresolved Munk layer.  As expected, one point in the layer is not sufficient to suppress the numerical mode in the velocity field.  We shall assimilate data of the fine resolution model in order to bring the solution of  the coarse resolution's model closer to fine resolution's one.

\begin{figure}
  \begin{center}
  \begin{minipage}[r]{1\textwidth} 
  \centerline{\includegraphics[angle=0,width=0.49\textwidth]{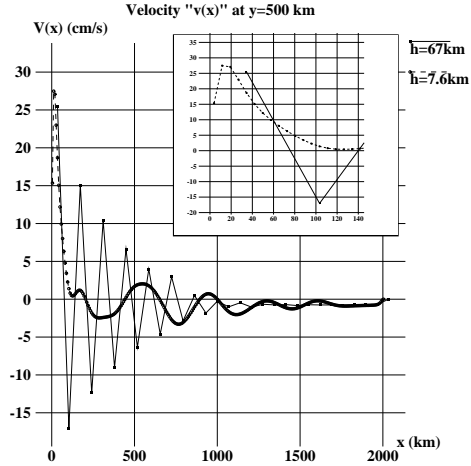}}
  \caption{ Profiles of the velocity $v(x,y)$ at $y=500$ km obtained on the coarse grid with $h=67km$ (solid line),  and fine grid with $h=7.6$ km (dashed line) }
  \end{minipage} 
  \end{center} 
\refstepcounter{fig}
\label{profile}
\end{figure}

The model on the fine grid has been spun up from the rest state during 3 years and integrated for the subsequent 3 years. From the result of this integration we have extracted values of all three variables at all grid points that belong to the coarse grid (as it has been noted, the grids have been chosen so, that all grid points of the coarse grid belong to the fine grid). This set   is used as artificial observations in the following experiments. 

So far the model is nonlinear with intrinsicly unstable solution, there is no hope to obtain close solutions in  long time model runs because any difference (even infinitesimal) between two models grows exponentially on these time scales. Consequently, we have to confine our study to the analysis of  relatively stable properties of the solution. So, we perform two experiments. The first one describes a short time evolution of the model's solution simulating the forecasting properties of the model.  The second one addresses the climatic averages of the model's solution which should also  represent more stable structures than particular trajectories (see \cite{sensbtp}, \cite{Dym2000} for example).  

In the first experiment we analyze the data assimilation that was used to identify optimal initial conditions of the model $\bar u_0, \bar v_0,\bar \eta_0$, optimal discretization of its operators near the boundary $\bar\alpha$  or both    applying minimization procedure and controlling different parameters as it is shown in  \rf{1stway},\rf{2ndway} and \rf{3rdway}. 
The final point of  spin-up of the high-resolution model was used as initial guess in  experiments that control the initial state of the low-resolution model.  Classical second order discretization of all operators near the boundary $ \alpha_{0}=0,\;  \alpha^{D}_{1}=-1/h,\;\alpha^{D}_{2}=1/h,\;  \alpha^{S}_{1}=\alpha^{S}_{2}=1/2 $ was assumed as initial guess in all experiments that control the discretization.   

In the  experiment that analyses the forecasting properties of the model, we assimilate ``observational data" during 5 days and we examine the evolution of the difference between the model's solution and observations during next 15 days.    

\begin{figure}
  \begin{center}
  \begin{minipage}[r]{1\textwidth} 
  \centerline{\includegraphics[angle=0,width=0.55\textwidth]{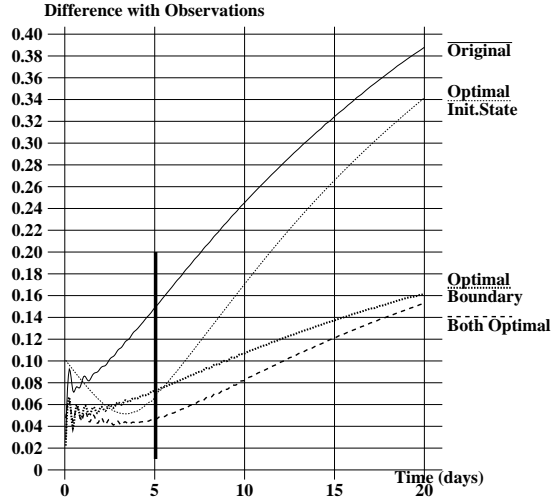}}
  \caption{ Evolution of the distance $\xi(t)$   \rf{xi} during and after assimilation.  }
  \end{minipage} 
  \end{center} 
\refstepcounter{fig}
\label{forecast-sq}
\end{figure}
 
 One can see in \rfg{forecast-sq} that when we control  initial conditions of the model, the control moves the initial state  (dashed line)  far from initial guess. The initial difference with observations becomes $\xi(0)=0.1$. However, at the end of the assimilation window the distance from observations  is reduced to $\xi(5 \mbox{ days})=0.07$. After the end of assimilation the  distance from observations increases rapidly. The line approaches the upper  solid line that corresponds to the solution of the model with no assimilation at all. This fact can easily be understood. The model's dynamics has not been modified by the assimilations procedure.  The model remains the same, possessing the same deficiencies. Consequently, it is not surprising that beyond assimilation window the solution tends to the attractor of the coarse resolution model.  The model starts to develop  spurious oscillations shown in \rfg{profile} moving away from the attractor of the high resolution model where these oscillations are absent. 
 
Controlling the discretization of model's operators, we modify the model. Assimilation, in this case, brings the model dynamics towards the  dynamics of the high resolution model (used to produce artificial observations).  This control does not modify the initial point ($\xi(0)=0$) but  the    distance $\xi(5 \mbox{ days})$ at the end of assimilation time  (dotted line in  \rfg{forecast-sq}) is almost equal to the distance obtained controlling the initial state of the model.  However, beyond the window, we see that the modified model's dynamics allows the solution to remain closer to observations. The difference $\xi$  on the twentieth day is more than  two times lower. If we suppose that we assimilate  data in the past 5 days in order to deliver a forecast for 15 days in the future, we see that controlling coefficients $\alpha$ provides  two times better result.

If we control both initial state and boundary discretization, we obtain  dashed line which is  similar to the dotted one. Indeed, the assimilation procedure was particularly concentrated on the control of coefficients $\alpha$ rather than on the control of the initial state, that remains rather close to the initial guess for this state  ($\xi(0)=0.01$) . In this experiment we have  lower distance $\xi$ at the end of the assimilation window, but almost the same $\xi$ at the end of the forecasting time.  

Thus, we see in this experiment that if the model's dynamics suffers from low resolution and other numerical errors,  better forecast is achieved by controlling the model's operators rather than initial conditions. 

If we consider a long time behavior of the model, we should analyze  climatic averages of the solution rather than the difference between particular trajectories.  Due to intrinsic instability any  trajectories diverge and the value of  $\xi$ becomes determined by the attractor size. If we assimilate data in order to control   initial state of the model, we can not  hope to improve its climatic averages because we do not modify the model's dynamics. No matter from which state the model starts, the same dynamics determines the same attractor and the same climatic (calculates over sufficiently long time interval) averages. In the same time, data assimilation performed with the purpose  to control the discretization of operators near the boundary, does modify the dynamics.  In this case, together with the short time behavior,  we can hope to improve the model's climate. 

To  study the modification of the climatic averages, we perform another experiment starting from the same initial guess,  controlling also both initial state and parameters $\alpha$, but with assimilation window $T=100$ days.  Such a large window is necessary to collect an observational  information about a number of physical processes that determine long-time model behavior.  Optimal initial point  $\bar u_0, \bar v_0,\bar \eta_0$ and optimal $\bar\alpha$ found in the experiment are used in the 1000 days model run that means 10 times the assimilation window. 

Averages of the original coarse resolution model suffer a lot from the numerical effects that are present due to insufficient resolution.  All fields contain spurious oscillations (see \rfg{profile}) due to  unresolved Munk layer. The length of the jet-stream in the middle of the square is about 400 km (800-1000 km in the high resolution model) and the variability of the model's solution  is very low. Eddy kinetic and eddy potential energies of the  high resolution model show approximately 20 times bigger amplitudes.  These fields are not shown in the paper. 

Optimal discretization of operators in the boundary region allows us to correct these numerical errors. Average sea surface elevation $\bar\eta$ and eddy potential energy $ \overline{\mbox{EPE}}$ 
\beq 
\bar\eta(x,y)=\fr{1}{T}\int_0^T \eta(x,y,t)dt ,\;\;\; \overline{\mbox{EPE}}=\fr{1}{T} \int_0^T (\eta(x,y,t)-\bar\eta(x,y))^2 dt 
\eeq
obtained in the "observational"  run of the high resolution model and in the run of the low resolution model with optimal boundary discretization  are  presented  in \rfg{aver-ssh} and \rfg{aver-epe}.   One can see in  that  both averages and variability are very similar. That means, optimal discretization modifies the climate of the coarse resolution model bringing it closer to the reference model.

\begin{figure}[h]
  \begin{center}
  \begin{minipage}[r]{1\textwidth} 
  \centerline{\includegraphics[angle=-90,width=0.45\textwidth]{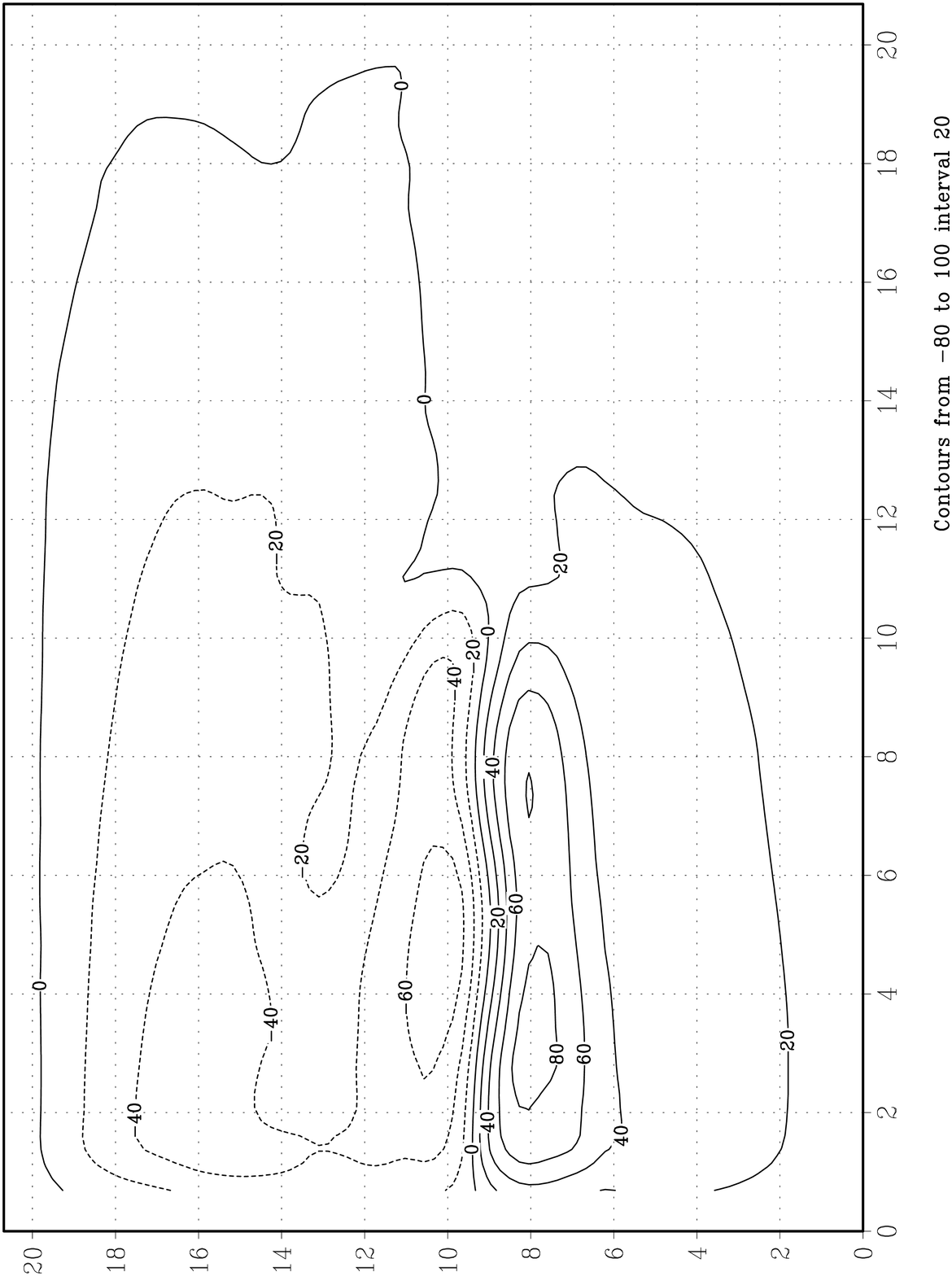}\includegraphics[angle=-90,width=0.45\textwidth]{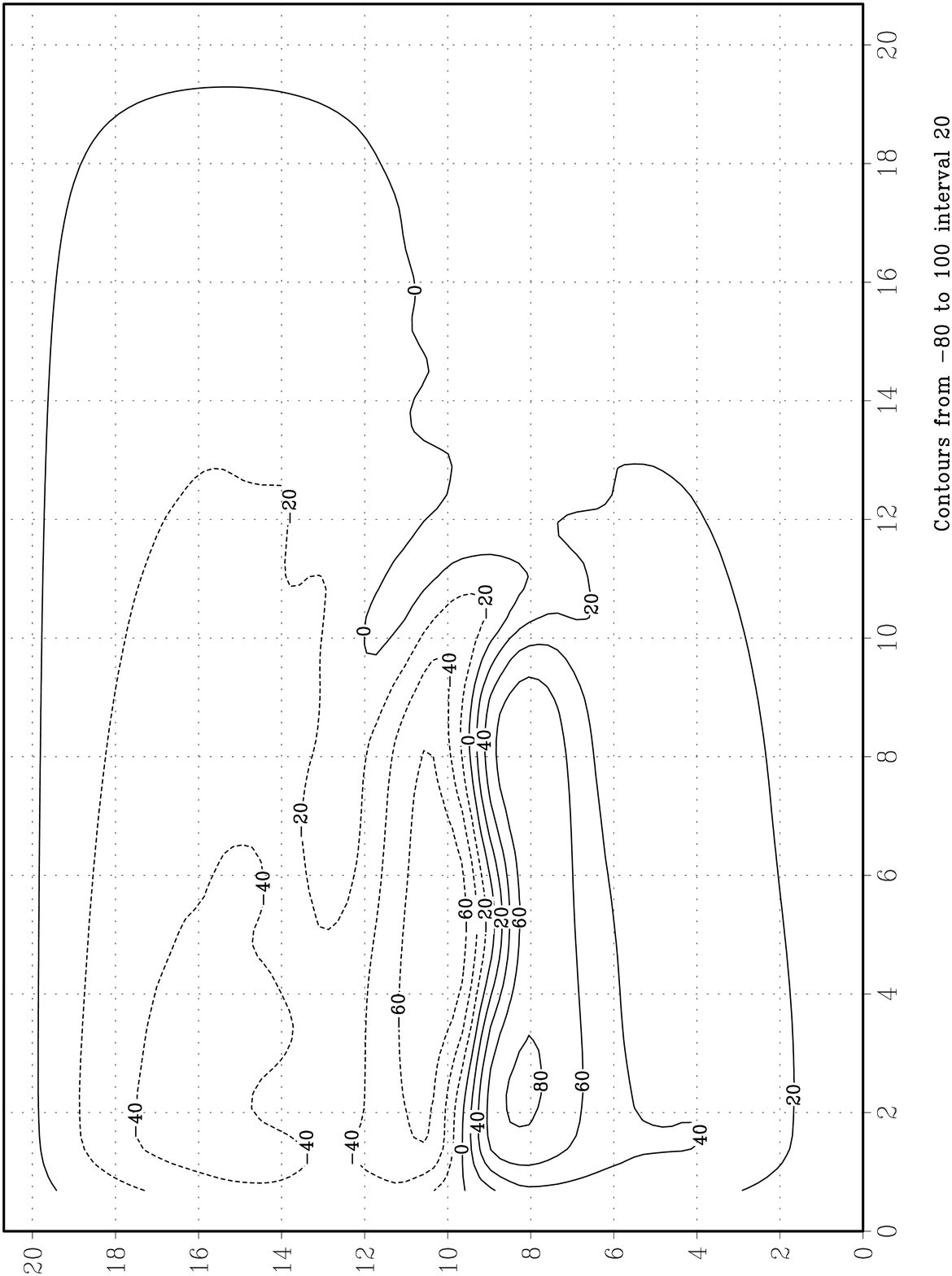}}
  \caption{ Thousand days average of the Sea Surface Elevation  obtained with the high resolution model (left) and with the low resolution model and optimal discretization (right).  }
  \end{minipage} 
  \end{center} 
\refstepcounter{fig}
\label{aver-ssh}
\end{figure}

\begin{figure}[h]
  \begin{center}
  \begin{minipage}[r]{1\textwidth} 
  \centerline{\includegraphics[angle=-90,width=0.45\textwidth]{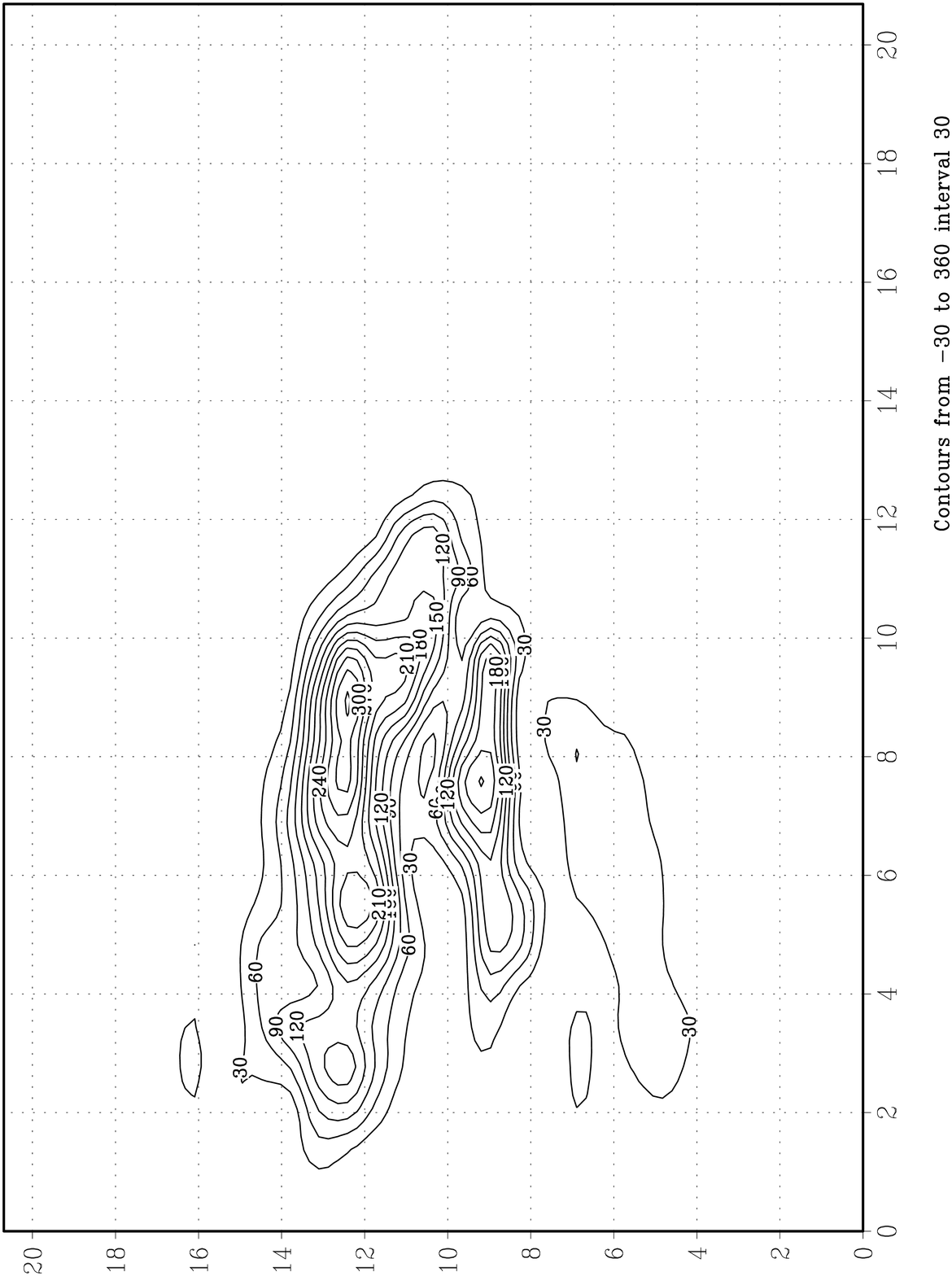}\includegraphics[angle=-90,width=0.45\textwidth]{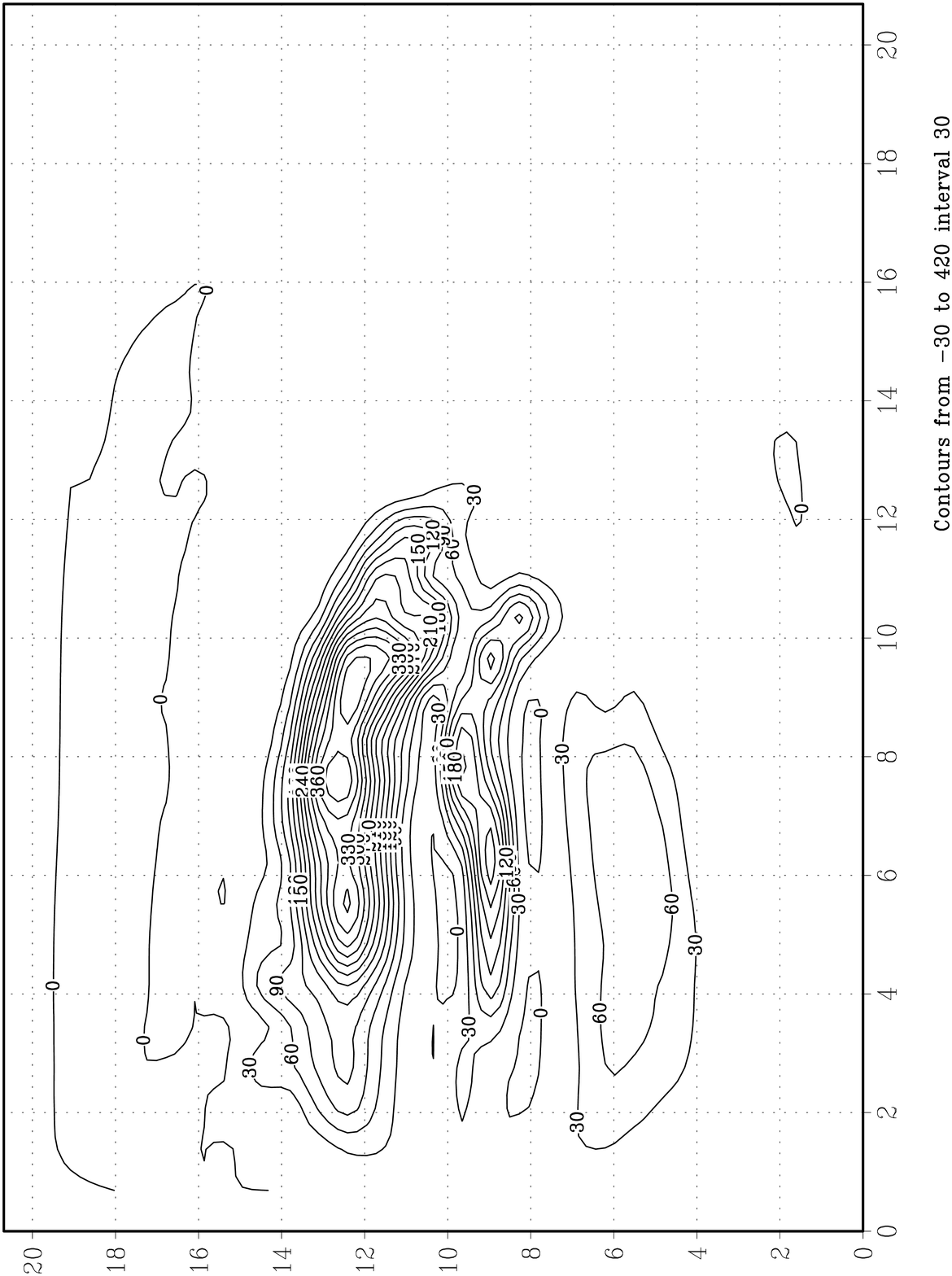}}
  \caption{ Thousand days average Eddy Potential Energy  obtained with the high resolution model (left) and with the low resolution model and optimal discretization (right).  }
  \end{minipage} 
  \end{center} 
\refstepcounter{fig}
\label{aver-epe}
\end{figure}

\section{Model of the Black Sea. }

In this section we use the same model, but all the parameters are defined to describe the upper layer circulation of  the Black sea. Configuration of the model  and  observational data have been kindly provided by Gennady Korotaev from the  Marine Hydrophysical Institute, National Academy of Sciences of
Ukraine, Sevastopol, Ukraine. This configuration is described in \cite{korot-model}. 

The model grid counts $141\tm 88$ nodes  that corresponds to the  grid box of dimension 7860 m and 6950 m in $x$ an $y$ directions respectively.   15 minutes  time step is
used for integration of the model.  The Coriolis parameter is equal to  $f_0= 10^{-4}s^{-1}$ and $\beta=2\tm 10^{-11} (ms)^{-1}$. Horizontal viscosity is taken as $\mu=200 m^2 s^{-1}$. Using a typical density difference,
$\rho_0 - \rho_1$ of $3.2 kg/m^3$ , and unperturbed layer thickness of $H_0 = 150 m$, the Rossby radius of deformation is estimated
at about 22 km. The grid therefore resolves the mesoscale processes reasonably well.

The model has been forced by the ECMWF wind stress data, available as daily averages for the years 1988 through
1999. 
Dynamical sea level reconstructed in \cite{korot-altim} was used as  observational data in this section. These data   have been collected in ERS-1 and TOPEX/Poseidon missions and preprocessed by the NASA Ocean Altimeter Pathfinder Project, Goddard Space Flight Center. Observational data are available from the 1st May 1992 until 1999. These data have been linearly interpolated to the model grid.

So far  the sea surface elevation is the only observational variable available in this experiment, we put $w_u=w_v=0$ in \rf{xi}. Consequently, the difference between the model's solution and observations is calculated taking into account the variable $\eta$  only.  

The behavior of the model  solution is not chaotic in this configuration. Variability in the model is generated directly by the variability of the wind stress on the surface.  Consequently, we can compare particular trajectories of the model on any time interval because their evolution is stable without exponential divergence. 

As it has been already noted, absence  of  observational data for the velocity fields brings   us to modify the cost function. We have to add the background term in the cost function in order to  require the velocity field to be sufficiently smooth. Otherwise, lack of information about velocity components in observational data would result in a spuriously irregular fields obtained in assimilation. To ensure necessary regularity of $u$ and $v$ we add the distance from the initial guess to the cost function \rf{costfn}. In order to emphasize the requirement of smoothness,  this distance is measured as an enstrophy of the difference between the initial guess and current state: 
\beq
\costfun_{smooth}=  \sum_{i,j}\biggl(\der{(v_{i,j}-v_{i,j}^0)}{x}-\der{(u_{i,j}-u_{i,j}^0)}{y}\biggr)^2  \label{costfn-smooth}
\eeq
where $u^0,\; v^0$ denote  velocity components of the initial guess of the minimization procedure. Of course, this term is only taken into account in the identification of the initial state of the model. 

Moreover, using real observational data requires to add at least one another term to the cost function. One can see in 
the Figure 2 in \cite{korot-altim},  spatially averaged  sea surface elevation of the Black sea exhibits a well distinguished seasonal cycle. That means the mass is not constant during a year, it decreases in autumn and increases in spring. Consequently, if we assimilate data during a short time (a season or less), we assimilate also the information about the  mass flux specific for this season. This flux can not be corrected later by the model because the discretization of operators near the boundary (that controls the mass evolution) is obtained once for all seasons.  The mass variation of the Black sea  reaches 25 centimeters of the sea surface elevation. Assimilating data  within one season may, consequently,  result in a persisting increasing or decreasing of the seal level of order of 50 cm  per year.   To avoid this spurious change of the total mass, we must either take the assimilation window of at least one year, or prescribe the mass conservation to the model's scheme. One year assimilation window is computationally expensive and is not justified by the model's physics. On the other hand, prescribed mass conservation removes just the sinusoidal seasonal variation, allowing us to keep all other processes and to choose any assimilation window we need. 

To correct the mass flux of the model, we add the following term to the cost function
\beq
\costfun_{mass}= \int\limits_0^T \biggl(\sum_{i,j}(\eta_{i,j}(t)-\eta_{i,j}(0))\biggr)^2 dt \label{costfn-mass}
\eeq
Similarly to \rf{costfn-smooth}, this term also ensures the regularity of the solution, but it is  taken into account when the assimilation is performed for  identification of the boundary parametrization. It can be noted here that  other terms may be added to the cost function in order to make a numerical scheme  energy and/or enstrophy conserving, but we do not use them in this paper. 

The total cost function in this section is composed of three parts: \rf{costfn}, \rf{costfn-smooth} and \rf{costfn-mass}:
\beq
\costfun_{total}= \costfun+\gamma_1\costfun_{smooth}+\gamma_2\costfun_{mass}\label{costfn-total}
\eeq
Coefficients $\gamma$ are introduced to weight the information that comes from observational data (with $\costfun$) and an a priori knowledge about mass conservation and regularity of the solution. 

This modification of the cost function results, of course, in additional terms in the gradient:
\beq
\nabla  \costfun_{total}= \nabla\costfun+ 2\gamma_1 \biggl( D_y^*D_y (u-u_0) + D_x^*D_x (v-v_0)\biggr) 
+2\gamma_2 \sum_{i,j} \biggl(\eta_{i,j}(t)-\eta_{i,j}(0)\biggr).
\eeq

The model is spun up  from the beginning of 1988 to May 1992 using the wind tension data on the surface. The state corresponding to the 1st of May 1992 12h GMT is used as the initial guess in the data assimilation procedure controlling initial conditions of the model.  The assimilation is performed   following the procedure \rf{1stway} with the assimilation window $T=1$ day and the regularization parameter $\gamma_1=0.04$.   Such a short window was chosen in order to get almost instantaneous state of the model to be used in further experiment as an initial state. 

This assimilation provides sufficiently smooth velocity fields (see \rfg{blk-initstate} left) that contain such a  specific features of the Black Sea circulation as Western and Eastern gyres, Batumi gyre, filament formation along the Caucasus coast and a formation of the Sebastopol eddy (see \cite{korot-model} for discussion of these features). Obtained sea surface elevation field (see \rfg{blk-initstate} right) is close to observational data that was used in the assimilation. The value of the difference $\xi$ \rf{xi} is less than 0.1, that means the actual average  difference between observed and reconstructed fields is approximately equal to  30 cm. 

\begin{figure}[h]
  \begin{center}
  \begin{minipage}[r]{1\textwidth} 
  \centerline{\includegraphics[angle=-90,width=0.49\textwidth]{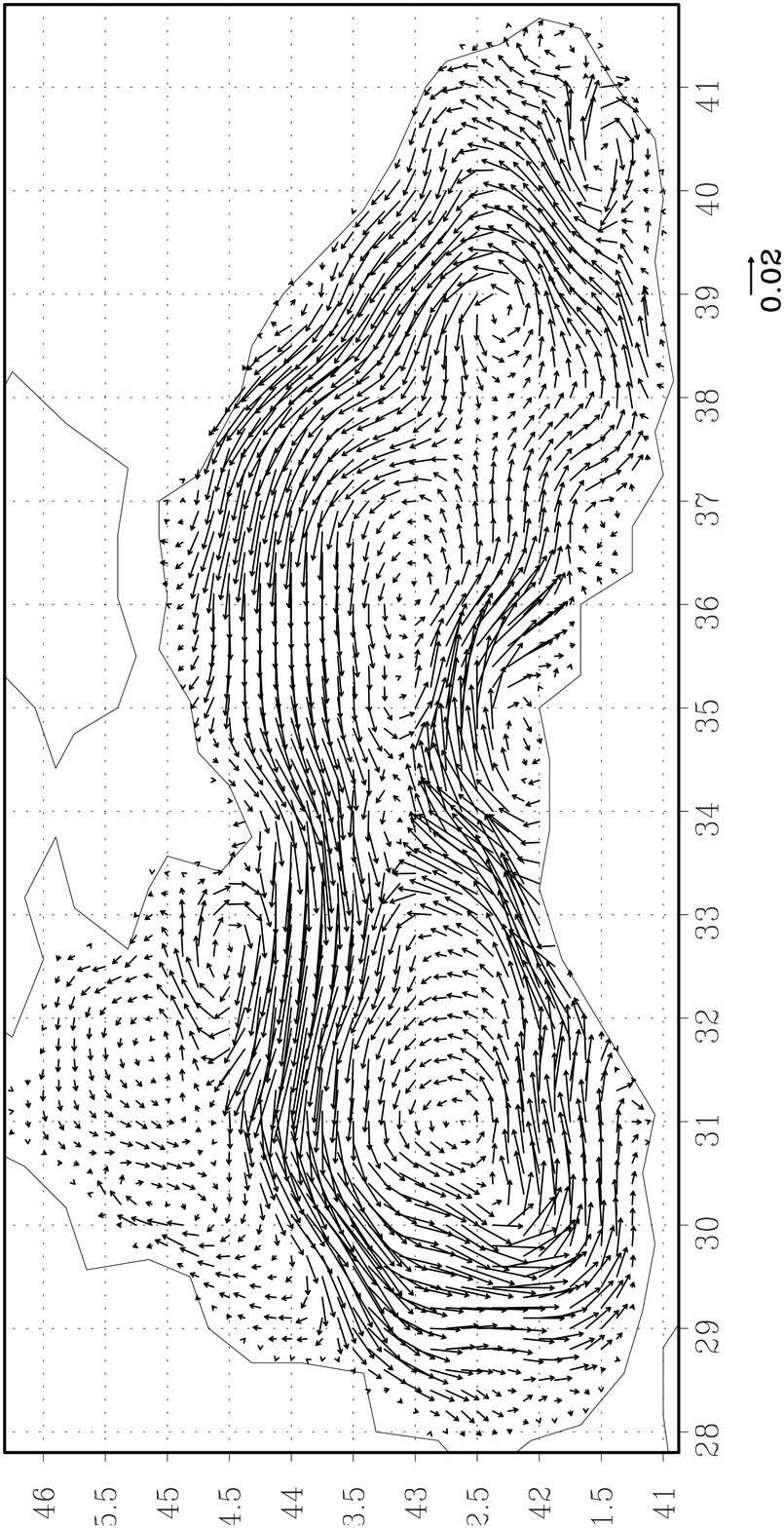}\includegraphics[angle=-90,width=0.49\textwidth]{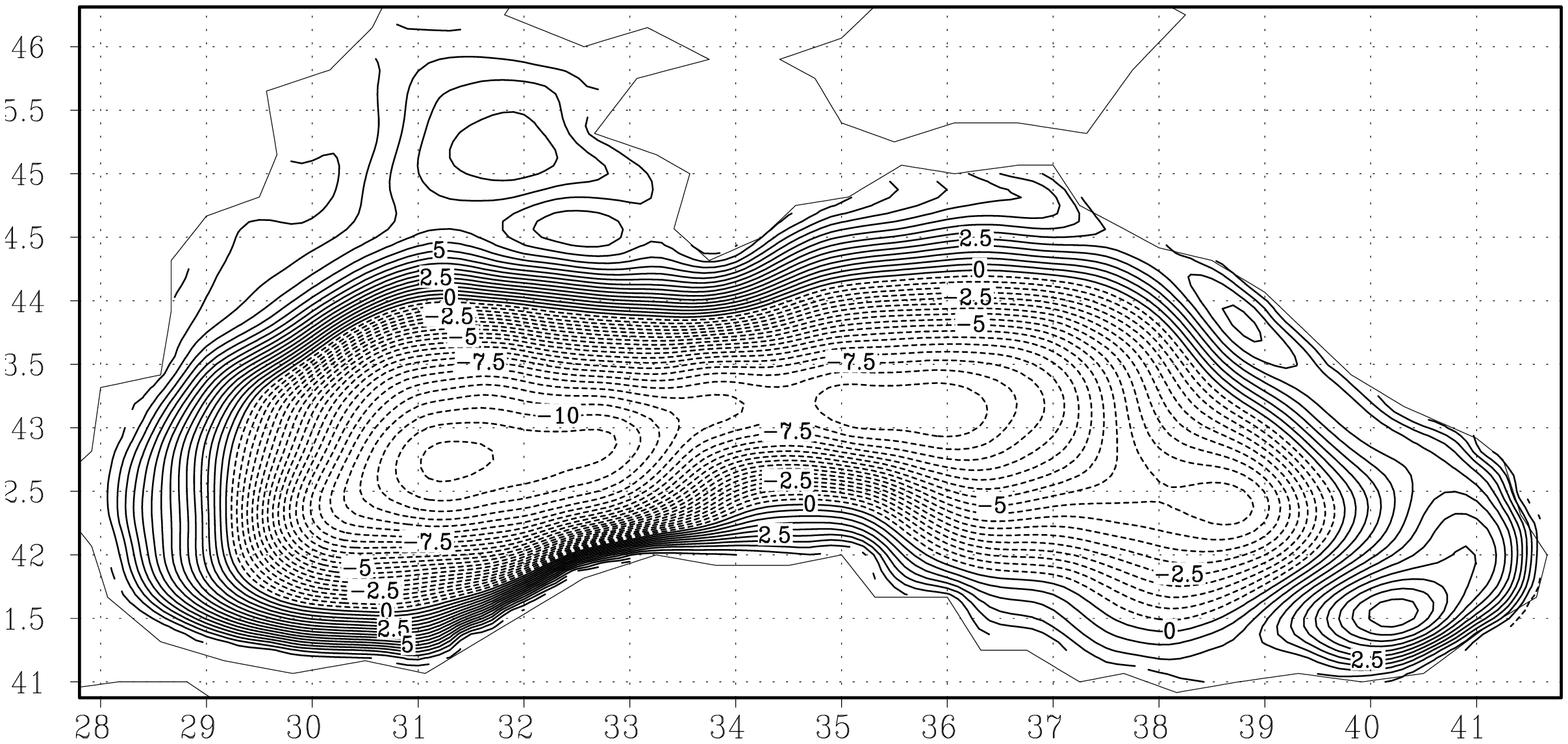}}
  \caption{ The model initial  state corresponding to the 1st of May 1992 obtained by data assimilation: velocity (left), sea surface elevation (right).   }
  \end{minipage} 
  \end{center} 
\refstepcounter{fig}
\label{blk-initstate}
\end{figure}

This initial state is used in data assimilation experiments that control the discretization of the model's operators near boundary.  Of course, short assimilation window is not appropriate to identify internal model's parameters. Observational data must contain information, and especially  about long time model behavior and must be capable to identify optimal discretization which is supposed to be constant in time. We have already   shown on the example of the total mass evolution that short assimilation windows may lead to  a wrong model behavior.   However, choosing a long assimilation window require much  computer time. 

In this paper we chose $T=50$ days window which is longer than synoptic time scales. The minimization of the cost function has been accompanied by the mass preserving correction \rf{costfn-mass} with $\gamma_2= 0.01$. 

As well as in the previous section, we perform 3 experiments controlling initial conditions, boundary parametrization and both of them. However, due to internal stability of the model solution, we can compare particular trajectories on long time intervals.  Evolution of the difference $\xi$ between the model's solution and observations is shown in \rfg{s21-blk}.    As we can see,  optimal initial point allows the solution to remain close to observations in  the assimilation window but not beyond the window. As soon as the assimilation ends, the solution goes rapidly toward the solution of the original model and becomes indistinguishable  from it after 200 days.  On the other hand, the solution of the model with optimally discretized operators remains always closer to the observational data than the solution of the original model. That means modified  model's dynamics allows the solution to approach observations. 

Comparing the computational cost of the data assimilation, we must acknowledge that controlling the boundary is more expensive. First, the adjoint model's run requires approximately double computer  time comparing with the adjoint model for  the identification of the initial state. And second,   the minimization process requires more iterations. In the experiment shown in  \rfg{s21-blk}, 5 iterations are necessary to minimize the cost function controlling   the initial state of the model, but we need at least 22 iterations to reach the same stopping criterion controlling $\alpha$.

\begin{figure}[h]
  \begin{center}
  \begin{minipage}[r]{1\textwidth} 
  \centerline{\includegraphics[angle=0,width=0.55\textwidth]{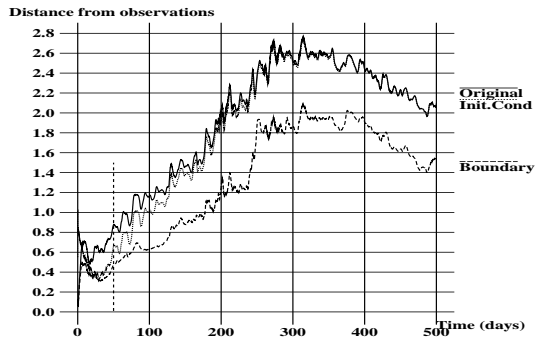}}
  \caption{ Evolution of the distance $\xi(t)$   \rf{xi} during  and after assimilation.   }
  \end{minipage} 
  \end{center} 
\refstepcounter{fig}
\label{s21-blk}
\end{figure}

As it has been discussed  in \cite{assimbc1}, it is difficult to distinguish principal  modifications that have been made in the numerical scheme by the data assimilation. Various kernels  are present in the space of $\alpha$ even in a very simple model like one-dimensional wave equation. These kernels make difficult  the analysis of the assimilation results because  optimal coefficients are not unique. Obtained  set of  $\alpha$ represent just one point in the kernel, while the same model behavior can be obtained with any other kernel point.  Consequently, in two similar assimilation experiments we can obtain very different sets of  optimal coefficients but almost the same model behavior.  

 In the present non-linear model  it is also difficult to analyze coefficients $\alpha$ directly. We have performed several experiments assimilating observational data during the same seasons with the same assimilation windows but in different years (1992, 1993, 1995). Assimilation results (not shown here) reveal very close values of the cost function obtained in the minimization procedure, very similar evolution of the difference $\xi$ in the assimilation window and beyond it (like shown in \rfg{s21-blk}), but very different sets of coefficients $\alpha$. As well as in experiments with simpler models (\cite{assimbc1},\cite{sw-lin}), this non-uniqueness of the optimal discretization coefficients  has also its roots in the kernel that exists  in the space of $\alpha$. Different points in the kernel correspond to different discretizations of the model's operators, that result in almost the same model's solutions. 
 
 Instead of analyzing  the set of obtained coefficients $\alpha$, we shall see the modification of the solution that this set generates, and namely the difference between  the velocity of the model with classical boundary (the distance of this solution from observations is plotted by the solid line in \rfg{s21-blk}) and the velocity with optimal boundary discretization (dashed line in  \rfg{s21-blk}).  This difference has been averaged in time over 200 days time interval in order to reveal persistent modifications of the flow produced by the optimal discretization. 
 
This average difference of the velocity is presented in \rfg{diff-vel}. We zoom  the Southern part of the Black sea because it is in this region  the difference shows the biggest values reaching 15 $\fr{cm}{s}$ while in the middle of the sea it rarely exceeds 1  $\fr{cm}{s}$. 

\begin{figure}[h]
  \begin{center}
  \begin{minipage}[r]{1\textwidth} 
  \centerline{\includegraphics[angle=-90,width=0.95\textwidth]{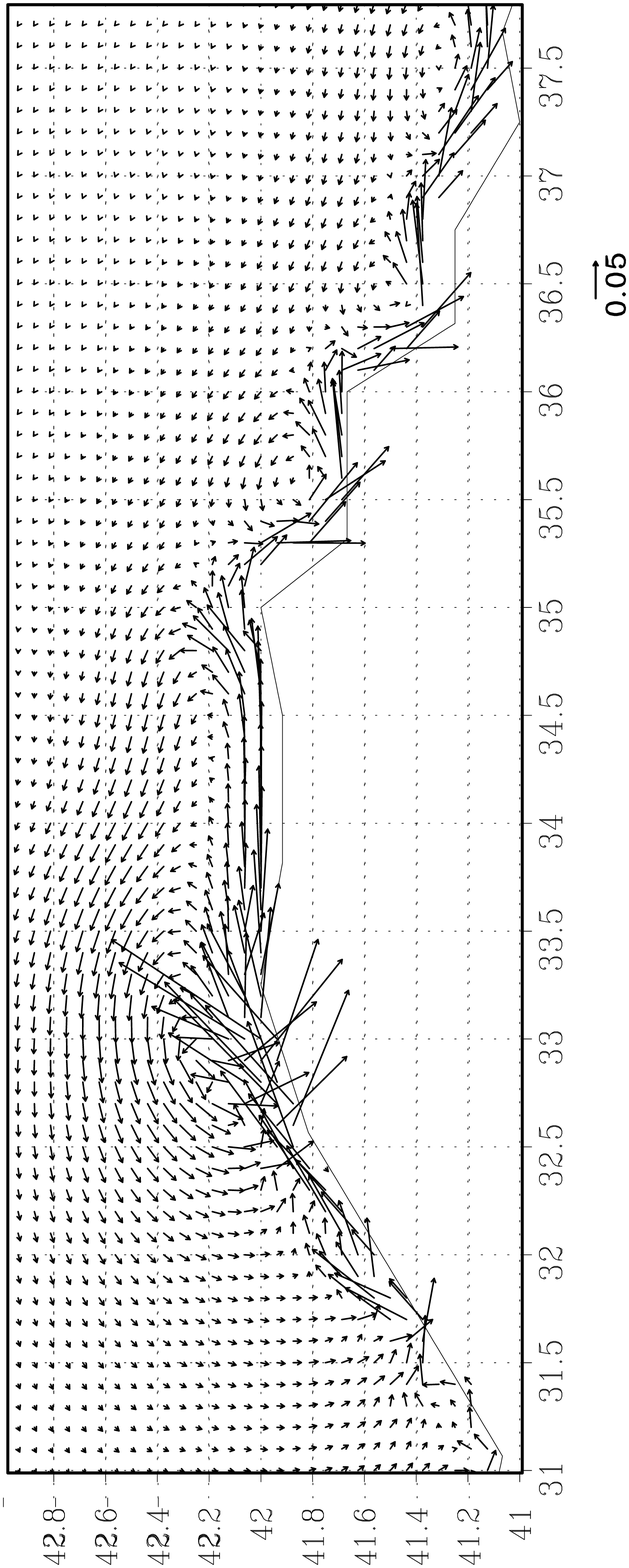}}
  \caption{Difference in the velocity field of solutions with classical and optimal discretizations.    }
  \end{minipage} 
  \end{center} 
\refstepcounter{fig}
\label{diff-vel}
\end{figure}

We can note several principal features of the flow that have been modified by   boundary conditions. First, we can see a strong  current on the boundary. The slip condition (vanishing tangential velocity) has been replaced by a permanent current along the boundary. Moreover, impermeability condition has also been modified. The flow is now  allowed to leave  the domain ensuring, however, the global mass balance.   One can see a strong persistent vortex centered at 
$42.2^\circ$N, $32.8^\circ$E which southern part crosses the boundary resulting in not only tangential but normal flux also.    Similar vortices with lower amplitude can also be seen in places where the boundary changes direction. Optimal discretization allows the flow to cross the boundary in places where the direction change is not smooth. 

Tangential velocity component is amplified in the direct vicinity of the boundary. In these nodes we see a strong eastward flow that was  forbidden by the boundary conditions in the classical formulation of the model. On the other hand, the eastward velocity is lower at nodes distanced by several grid cells from the boundary. At these nodes we see westward flow in the difference of  the optimally discretized and classical models. In fact, the control of the discretization of operators near the boundary results in the same phenomenon as we have seen above in the experiments with the square box showing the Munk boundary layer \rfg{profile}.  A strong tangential current in the boundary layer is moved on the boundary  allowing more optimal  representation of a thin current on a coarse grid that brings the  model solution towards observations.

\section{Conclusion}

This paper is an extension of the study presented in \cite{sw-lin}. Both papers discuss the variational data assimilation procedure applied for identification of the optimal parametrization of  derivatives and interpolation operators near the boundary, but now we work with nonlinear  models assimilating both artificial and real observational data. 

Contrary to linear models, even assimilating artificially generated data in twin experiments, we can not obtain a solution  that is indistinguishable from assimilated data.  Due to intrinsic instability, even initially close solutions of a non-linear chaotic model diverge. Hence, any difference between models in twin experiments grows with time and prevent the solutions to remain close.  
Thus, dealing with nonlinear models, we can control the discretization  near the boundary. This allows us to correct the numerical errors due to insufficient resolution (see \rfg{profile}), to bring the model solution closer to observations (see \rfg{forecast-sq}, \rfg{s21-blk})  and to improve statistical averages of a chaotic model's solution (see \rfg{aver-ssh}, \rfg{aver-epe}).

Controlling the discretization of operators of the model possesses another advantage.   The solution with optimal discretization  remains closer to observational data after the assimilation end than the solution with optimal initial conditions. This fact can be observed both in the  square box and in the Black sea. Starting from the optimal initial point, the trajectory remains close to observations in the assimilation window, but  the distance with observations increases rapidly beyond the window. The solution with optimal discretization remains closer to observations even after the end of assimilation time. That means, observational data in the past are more efficiently used  for forecasting if we assimilate them to control internal model's parameters than to control the initial point.

{\bf Acknowledgments. } Author thanks  Gennady Korotaev from   Marine Hydrophysical Institute, National Academy of Sciences of
Ukraine for providing the model parameters and data for the upper layer model of the Black Sea. 
 

\begin{thebibliography}{}

\end{thebibliography}


\begin{thebibliography}{10}
\providecommand{\url}[1]{\texttt{#1}}
\providecommand{\urlprefix}{URL }
\expandafter\ifx\csname urlstyle\endcsname\relax
  \providecommand{\doi}[1]{doi:\discretionary{}{}{}#1}\else
  \providecommand{\doi}{doi:\discretionary{}{}{}\begingroup
  \urlstyle{rm}\Url}\fi

\bibitem{Ledimet82}
Le~Dimet FX. A general formalism of variational analysis. \emph{Technical
  {R}eport OK 73091}, CIMMS report, Normann 1982.

\bibitem{ldt86}
Le~Dimet FX, Talagrand O. Variational algorithm for analysis and assimilation
  of meteorological observations. theoretical aspects. \emph{Tellus}  1986;
  \textbf{38A}:97--110.

\bibitem{Lions68}
Lions JL. \emph{Contr\^ole optimal de syst\`emes gouvern\'es pas des
  \'equations aux d\'eriv\'ees parielles.} Dunod, 1968.

\bibitem{Marchuk75}
Marchuk G. Formulation of theory of perturbations for complicated models.
  \emph{Appl. Math. Optimization}  1975; \textbf{2}:1--33.

\bibitem{LoschWunsch}
Losch M, Wunsch C. Bottom topography as a control variable in an ocean model.
  \emph{J. Atmospheric and Oceanic Technology}  2003; \textbf{20}:1685--1696.

\bibitem{Heemink}
Heemink AW, Mouthaana EEA, Roesta MRT, Vollebregta EAH, Robaczewskab KB,
  Verlaanb M. Inverse 3d shallow water flow modelling of the continental shelf.
  \emph{Continental Shelf Research}  2002; \textbf{22}:465--484.

\bibitem{assimtopo}
Kazantsev E. Identification of optimal topography by variational data
  assimilation. \emph{J. Phys. Math.}  2009; \textbf{1}:1--23.

\bibitem{shulman97}
Shulman I. Local data assimilation in specification of open boundary
  conditions. \emph{J. of Atmospheric and Oceanic Technology}  1997;
  \textbf{14}:1409--1419.

\bibitem{shulman98}
Shulman I, Lewis JK, Blumberg AF, Kim BN. Optimized boundary conditions and
  data assimilation with application to the m2 tide in the yellow sea. \emph{J.
  of Atmospheric and Oceanic Technology}  1998; \textbf{15}(4):1066--1071.

\bibitem{Taillandier}
Taillandier V, Echevin V, Mortier L, Devenon JL. Controlling boundary
  conditions with a four-dimensional variational data-assimilation method in a
  non-stratified open coastal model. \emph{Ocean Dynamics}  2004;
  \textbf{54}(2):284--298.

\bibitem{Brummelhuis}
ten Brummelhuis PGJ, Heemink AW, Van~Den Boogaard HFP. Identification of
  shallow sea models. \emph{International Journal for Numerical Methods in
  Fluids}  1993; \textbf{17}:637--665.

\bibitem{zou}
Zou X, Navon I, Le~Dimet FX. An optimal nudging data assimilation scheme using
  parameter estimation. \emph{Q. J. of Roy. Met. Soc.}  1992;
  \textbf{118}:1163--1186.

\bibitem{panchang}
Panchang V, O'Brien J. \emph{On the Determination of Hydraulic Model Parameters
  Using the Strong Constraint Formulation Modeling Marine Systems}, vol.~1,
  chap.~1. CRC Press Inc, 1988; 5--18.

\bibitem{chertok}
Chertok  DL,  Lardner RW. Variational data assimilation for a nonlinear hydraulic model.
  \emph{Applied mathematical modelling}  1996; \textbf{20}(9):675--682.

\bibitem{VerronBlayo}
Verron J, Blayo E. The no-slip condition and separation of western boundary
  currents. \emph{J. Phys. Oc.}  1996; \textbf{26}(9):1938--1951.

\bibitem{ChenLin}
Chen H, Lin S, Wang H, Fang LC. Estimation of two-sided boundary conditions for
  twodimensional inverse heat conduction problems. \emph{Int. J. Heat Mass
  Transfer}  2002; \textbf{45}:15--43.

\bibitem{GillijnsDeMoor}
Gillijns S, Moor BD. Joint state and boundary condition estimation in linear
  data assimilation using basis function expansion. \emph{Modelling,
  Identification, and Control}, Bruzzone L (ed.), 2007.

\bibitem{fxld-mo}
Le~Dimet FX, Ouberdous M. Retrieval of balanced fields: an optimal control
  method. \emph{Tellus A}  1993; \textbf{45}(5):449--461.

\bibitem{assimbc1}
Kazantsev E. Identification of an optimal boundary approximation by variational
  data assimilation. \emph{J.Comp.Phys.}  2010; \textbf{229}:256--275.

\bibitem{sw-lin}
Kazantsev E. Optimal boundary discretisation by variational data assimilation.
  \emph{Int. J. for Numerical Methods in Fluids}  2010;  n/a. doi: 10.1002/fld.2240.

\bibitem{Leredde}
Leredde Y, Lellouche JM, Devenon JL, Dekeyser I. On initial, boundary
  conditions and viscosity coefficient control for burgers' equation.
  \emph{International Journal for Numerical Methods in Fluids}  1998;
  \textbf{28}(1):113--128.

\bibitem{Lellouche}
Lellouche J, Devenon J, Dekeyser I. Boundary control on burgers equation. a
  numerical approach. \emph{Comput. Math. Appl.}  1994; \textbf{28}:33--44.

\bibitem{Gill}
Gill A. \emph{Atmosphere-Ocean Dynamics}. Academic Press, 1982.

\bibitem{Pedlosky}
Pedlosky J. \emph{Geophysical Fluid Dynamics}. 2d ed., Springer-Verlag, 1987.

\bibitem{AL77}
Arakawa A, Lamb V. \emph{Computational Design of the Basic Dynamical Processes
  of the UCLA General Circulation Model}, vol.~17, chap. Methods in
  Computational Physics. Academic Press, 1977; 174--267.

\bibitem{AL81}
Arakawa A, Lamb V. A potential enstrophy and energy conserving scheme for the
  shallow water equations. \emph{Mon. Wea. Rev.}  1981; \textbf{109}:18--36.

\bibitem{UnifNotat}
Ide K, Courtier P, Ghil M, Lorenc A. Unified notation for data assimilation:
  Operational, sequential and variational. \emph{J. of the Met. Soc. of Japan}
  1997; \textbf{75}(1B):181--189.

\bibitem{lemarechal}
Gilbert J, Lemarechal C. Some numerical experiments with variable storage
  quasi-newton algorithms. \emph{Mathematical programming}  1989;
  \textbf{45}:407--435.

\bibitem{LPV}
Le~Provost C, Verron J. Wind-driven ocean circulation transition to barotropic
  instability. \emph{Dyn.Atmos.Oceans}  1987; \textbf{11}:175--201.

\bibitem{simmonet2}
Simmonet E, Ghil M, Ide K, Temam R, Wang S. Low-frequency variability in
  shallow-water models of the wind-driven ocean circulation. \emph{J. Phys.
  Oc.}  2003; \textbf{33}:729--752.

\bibitem{Munk}
Munk WH. On the wind-driven ocean circulation. \emph{Journal of Meteorology}
  1950; \textbf{7}:3--29.

\bibitem{Bryan75}
Bryan K, Manabe S, Pacanowski R. A global ocean-atmosphere climate model. part
  ii. the oceanic circulation. \emph{J. Phys. Oc.}  1975; \textbf{5}(1):30--46.

\bibitem{Griffies}
Griffies SM, Pacanowski RC, Hallberg RW. Spurious diapycnal mixing associated
  with advection in a z-coordinate ocean model. \emph{Mon. Wea. Rev.}
  2000; \textbf{128}(3):538--564.

\bibitem{sensbtp}
Kazantsev E. Sensitivity of the attractor of the barotropic ocean model to
  external influences: Approach by unstable periodic orbits. \emph{Nonlinear
  processes in Geophysics}  2001; \textbf{8}:281--300.

\bibitem{Dym2000}
Dymnikov VP. Modelling of the climatic system response to small external
  forcing. \emph{Russ. J. Numer. Anal. Math. Modelling}  2000;
  \textbf{15}(1):15--28.

\bibitem{korot-model}
Korotaev G, Oguz T, Nikiforov A, Koblinsky C. Seasonal, interannual, and
  mesoscale variability of the black sea upper layer circulation derived from
  altimeter data. \emph{JGR}  2003; \textbf{108}:3122.

\bibitem{korot-altim}
Korotaev GK, Saenko OA, Koblinsky CJ. Satellite altimetry observations of the
  black sea level. \emph{JGR}  2001; \textbf{106}:917--934.

\end{thebibliography}

\mkpicstoend 

\end{document}